\documentclass[a4paper,11pt]{article}
\usepackage{amsfonts,latexsym,epsfig}

\newlength{\dinwidth}
\newlength{\dinmargin}
\newtheorem{theorem}{Theorem}

\newtheorem{remark}{Remark}
\newtheorem{lemma}{Lemma}

\def\s{\sigma}
\def\L{{\cal L}}
\def\Z{{\mathbb Z}}

\newcommand{\Gl}{\lambda}

\newcommand{\cB}{{\cal B}}

\newcommand{\cH}{{\cal H}}

\newcommand{\hcH}{{\widehat{\cH}}}
\def\la{\label}

\def\p{\partial}
\def\l{\lambda}
\def\f{\frac}

\newcommand{\ft}[2]{{\textstyle {\frac{#1}{#2}} }}

\newcommand{\tr}{{\rm tr }}

\newcommand{\I}{{\rm i}}

\newcommand{\qed}{\begin{flushright}$\Box$\end{flushright}}

\newcommand{\be}{\begin{equation}}
\newcommand{\ee}{\end{equation}}
\newcommand{\ben}{\begin{displaymath}}
\newcommand{\een}{\end{displaymath}}
\newcommand{\ba}{\begin{eqnarray}}
\newcommand{\ea}{\end{eqnarray}}
\newcommand{\bea}{\begin{eqnarray}}
\newcommand{\eea}{\end{eqnarray}}

\newcommand{\non}{\nonumber\\}

\newcommand{\mathon}{\mathversion{bold}}
\newcommand{\mathoff}{\mathversion{normal}}

\newcommand{\Ref}[1]{(\ref{#1})}

\makeatletter
\@addtoreset{equation}{section}
\makeatother

\newcommand{\HH}[1]{\hcH_{#1}}
\newcommand{\LH}[2]{L_{#1}\hcH_{#2}}
\newcommand{\LLH}[3]{L_{#1}L_{#2}\hcH_{#3}}
\newcommand{\LLLH}[4]{L_{#1}L_{#2}L_{#3}\hcH_{#4}}

\def\p{\partial}

\begin{document}

\begin{flushright}\small 
\end{flushright}
%
\vskip 10mm
\begin{center}
\mathon
  {\LARGE {\bf Generalization of Okamoto's equation \\[1ex] 
  to arbitrary  $2\times 2$ Schlesinger systems}}
\mathoff
\end{center}
\vskip 5mm

\begin{center}
{\large{\bf D.~Korotkin${}^{\rm 1}$ and H.~Samtleben${}^{\rm 2}$}}
  \\[6mm]

${}^1$ 
Department of Mathematics and Statistics, \\
Concordia University, 7141 Sherbrooke West, \\
Montreal H4B 1R6, Quebec, Canada
\\[2mm]
${}^2$ Universit\'e de Lyon, Laboratoire de Physique, \\
Ecole Normale Sup\'erieure de Lyon,\\ 
46, all\'ee d'Italie, F-69364 Lyon CEDEX 07, France \\[3mm]
{{\tt korotkin@mathstat.concordia.ca}\,,\,{\tt henning.samtleben@ens-lyon.fr}}
\end{center}

\vskip .2in

\begin{center} {\bf Abstract } 
\end{center}
\begin{quotation}\noindent
The $2\times 2$ Schlesinger system for the case of four regular
singularities is equivalent to the Painlev\'e VI equation. The Painlev\'e VI
equation can in turn be rewritten in the symmetric form of
Okamoto's equation; the dependent variable in Okamoto's form of the
PVI equation is the (slightly transformed) logarithmic derivative of
the Jimbo-Miwa tau-function of the Schlesinger system. The goal of
this note is twofold. First, we find a symmetric uniform formulation of an arbitrary Schlesinger 
system with regular singularities in terms of appropriately defined Virasoro generators.
Second, we find  analogues of Okamoto's equation for the case of
the $2\times2 $ Schlesinger  system with an arbitrary number of poles. 
A new set of scalar equations for the logarithmic derivatives of the Jimbo-Miwa
tau-function is derived in terms of generators of the Virasoro algebra; 
these generators are expressed 
in terms of derivatives with respect to singularities of the
Schlesinger system. 
\end{quotation}

\bigskip
\bigskip
\bigskip

\section{Introduction}


The Schlesinger system is the following non-autonomous system of differential
 equations for $N$ unknown matrices $A_j\in \mathfrak{sl}(M)$ depending
on $N$ variables $\{\Gl_j\}$:
\be
\frac{\partial A_j}{\partial\Gl_i}=
\frac{[A_j,A_i]}{\Gl_j-\Gl_i}\;,\quad i\neq j\;,\qquad
\frac{\partial A_j}{\partial\Gl_j}=
-\sum_{i\neq j}
\frac{[A_j,A_i]}{\Gl_j-\Gl_i}\;.
\label{sch}
\ee
The system (\ref{sch}) determines isomonodromic deformations of a solution of matrix ODE with meromorphic coefficients 
\bea
 \f{\p\Psi}{\p\Gl}= A(\lambda)\,\Psi \equiv 
 \sum_{j=1}^N\f{A_j}{\Gl-\Gl_j}\,\Psi\;.
 \eea
The  solution of this system normalized at a fixed  point $\Gl_0$ by $\Psi(\Gl_0)=I$ solves a
matrix Riemann-Hilbert problem with some monodromy matrices around the singularities $\Gl_j$.

The Schlesinger equations were discovered almost 100 years ago \cite{Schlesinger}; 
however, they continue to play a key role
 in many areas of mathematical physics: the theory of
random matrices, integrable systems, theory of Frobenius manifolds, etc.\,.
The system (\ref{sch}) is a non-autonomous hamiltonian system with respect to the Poisson bracket

\be
\{A_j^a,\;A_k^b\}=\delta_{jk}\,f^{ab}{}_c\,A_j^c\;,
\la{PoissonA}
\ee
where $f^{ab}{}_c$ are structure constants of $\mathfrak{sl}(M)$; $\delta_{jk}$ is the Kronecker symbol.
Obviously, the traces $\tr A_j^n$ are integrals of the Schlesinger system for any value of $n$. 
The commuting Hamiltonians defining evolution with respect to the times $\lambda_j$ are given by
\be
H_j= \frac1{4\pi\I}\oint_{\Gl_i} \tr A^2(\Gl)\,d\Gl
\equiv\f{1}{2}\sum_{k\neq j}\f{\tr A_j A_k}{\Gl_k-\Gl_j}
\;.
\la{hamintr}
\ee
The generating function $\tau_{\rm JM}(\{\Gl_j\})$ of the hamiltonians $H_j$, defined by 
\be
\f{\p}{\p\Gl_j}\log \tau_{\rm JM}= H_j\;,
\la{deftau}
\ee
was introduced by Jimbo, Miwa and their co-authors  \cite{Jimbo80,Jimbo81}; 
it is called the $\tau$-function of the Schlesinger system. 
The $\tau$-function plays a key role in the theory of the Schlesinger equations; in particular, the
divisor of zeros of the $\tau$-function coincides with the divisor of singularities of the solution of
the Schlesinger system; on the same divisor the underlying Riemann-Hilbert problem looses its solvability.

In the simplest non-trivial case when the matrix dimension
 equals  $M=2$ and the number of singularities equals $N=4$, the Schlesinger system
can equivalently be rewritten as a single scalar differential equation of order two
 ---
the Painlev\'e VI equation
\bea
  \frac{d^2y}{dt^2}&=&\frac 12\left(\frac 1y+\frac 1{y-1}+\frac 1{y-t}\right)
\left(\frac{dy}{dt}\right)^2-\left(\frac 1t+\frac 1{t-1}+\frac 1{y-t}\right)
\frac{dy}{dt}
\nonumber\\[.5ex]
&&{}+
\frac{y(y-1)(y-t)}{t^2(t-1)^2}\left(\alpha+\beta\frac t{y^2}+
\gamma\frac{t-1}{(y-1)^2}+\delta\frac{t(t-1)}{(y-t)^2}\right)\;,
  \label{P6}
\eea
where $t$ is the cross-ratio 
of the four singularities $\Gl_1,\dots,\Gl_4$, and $y$ is the position of 
a zero of the upper right corner element of the matrix $\sum_{k=1}^4 \f{A_k}{\Gl-\Gl_k}$.
Let us denote the eigenvalues of the matrices $A_j$ by $\alpha_j/2$ and $-\alpha_j/2$.
Then the constants $\alpha,\beta,\gamma$ and $\delta$ from the Painlev\'e VI equation (\ref{P6}) are related to
the constants $\alpha_j$ as follows:
\be
\alpha= \f{(\alpha_1-1)^2}{2} \;,\hskip0.5cm
\beta=-\f{\alpha_2^2}{2} \;,\hskip0.5cm
\gamma=\f{\alpha_3^2}{2} \;,\hskip0.5cm
\delta= \f{1}{2}-\f{\alpha_4^2}{2}\;.
\ee

It was further observed by Okamoto \cite{Okamoto0,Okamoto}, 
that the Painlev\'e VI equation
(and, therefore, the original $2\times 2$ Schlesinger system with
four singularities) can be rewritten alternatively 
in a simple form in terms of  the so-called auxiliary hamiltonian  function $h(t)$. 
To define this function we need to introduce first four  
constants $b_j$, which are expressed in terms of the eigenvalues of the matrices $A_j$ as
follows:
$$
b_1=\f{1}{2}(\alpha_2+\alpha_3)\;,\hskip0.6cm
b_2=\f{1}{2}(\alpha_2-\alpha_3)\;,
$$
\be
b_3=\f{1}{2}(\alpha_4+\alpha_1)\;,\hskip0.6cm
b_4=\f{1}{2}(\alpha_4-\alpha_1)\;.
\ee
The auxiliary hamiltonian function $h(t)$ is defined in terms of solution $y$ of equation (\ref{P6}) and the
constants $b_j$ as follows:
\bea
h&=& y(y-1)(y-t) \left(\f{dy}{dt}\right)^2 
\nonumber\\
&&{}
-\{(b_1+b_2)(y-1)(y-t)+(b_1-b_2)y(y-t)+(b_3+b_4)y(y-1)\}\f{dy}{dt} 
\nonumber\\
&&{}
+\left\{\f{1}{4}(2b_1+b_3+b_4)^2-\f{1}{4}(b_3-b_4)^2 \right\}(y-t)+\sigma_2'[b]t-\f{1}{2}\sigma_2[b]
\;,
\eea
where
\be 
\sigma_2'[b]:=b_1 b_3+b_1 b_4+ b_3 b_4\;,\hskip0.7cm
\sigma_2[b]:=\sum_{j,k=1\;\;j<k}^4 b_j b_k\;.
\ee
In terms of 
 the function $h$, the 
Painlev\' e equation (\ref{P6}) can be represented in 
a remarkably symmetric form as follows:
\bea
\f{d h}{dt}\left[t(1-t)\f{d^2 h}{d t^2}\right]^2+
\left[\f{d h}{dt}\left\{2h-(2t-1)\f{d h}{dt}\right\}+b_1 b_2 b_3 b_4\right]^2
-\prod_{k=1}^4\left(\f{d h}{dt} +b_k^2\right)&=&0
\;.\;\;\;\;\;\;
\la{Okameq}
\eea

Okamoto's form (\ref{Okameq}) of the Painlev\' e VI equation 
turned out to be extremely fruitful for 
establishing the hidden symmetries
of the equation (the so-called Okamoto symmetries). These symmetries look
very simple in terms of the auxiliary 
hamiltonian function $h$, but are highly non-trivial on the level of 
the solution $y$ of the
Painlev\' e VI equation, the corresponding monodromy group and the solution of 
the associated fuchsian system 
\cite{Dubrovin,Mazz}.

The goal of this paper is twofold. First, we show how to rewrite the Schlesinger system 
in an arbitrary matrix dimension in a symmetric universal form. Second, we use this symmetric form to find natural analogues of 
 the Okamoto equation (\ref{Okameq})
for  $2\times 2$ Schlesinger systems
with an arbitrary number of simple poles. Our approach is similar to the 
approach used by J.~Harnad to derive analogues of the Okamoto equation 
for Schlesinger systems corresponding to higher order poles 
(non-fuchsian systems)~\cite{Harnad}.

Namely, introducing the following differential operators (which satisfy the commutation relations of
the Virasoro algebra):
$$
L_m:=\sum_{j=1}^N \lambda_j^{m+1}\frac{\p}{\p\lambda_j}\;,\hskip0.7cm m= -1,0,1,\dots
\;,
$$
and the following dependent variables:
$$
\cB_n := \sum_j \Gl_j^{n}\,A_j\equiv {\rm res}|_{\Gl=\infty}\{ \Gl^{n} A(\Gl)\}\;,\hskip0.7cm n= 0,1,\dots\;,
$$
one can show that the Schlesinger system (\ref{sch}) implies 
\be
L_m \cB_n =
\sum_{k=1}^{n-1} \left[\cB_k,\cB_{m+n-k} \right] + n \cB_{m+n} 
\;,
\la{symint}
\ee
for all $n\geq 0$ and $m\geq -1$.  The infinite set of equations (\ref{symint}) is of course 
dependent for any given $N$.
To derive the original Schlesinger system 
(\ref{sch}) from (\ref{symint}) it is sufficient to take the set of equations (\ref{symint}) for
$n\leq N$ and $m\leq N$. The advantage of the system (\ref{symint}) is in its universality:
its form is independent of the number of the poles; the positions of the poles enter only the definition
of the differential operators $L_m$.

Consider now the case of $2\times 2$ matrices. To formulate the analog of the Okamoto equation for
the case of an arbitrary number of poles we introduce the following ``hamiltonians'':
$$
\hcH_m := -\ft14 \sum_{k=0}^m \tr\, \cB_k\,\cB_{m-k} \;,
$$
which can be viewed as symmetrised analogues of the Hamiltonians (\ref{hamintr}); they coincide with
$L_m \log\tau_{JM}$ up to an elementary transformation.
The simplest equation satisfied by $\hcH_m$ in the case of $2\times 2$ system is given by
\bea
\ft18\left(\LLH222 +2 \LH33 - 5 \LH42 -2 \HH6 \right)^2 
&=&
(\LH33-\LH42-\HH6) \, \Big( (\HH3)^2 +4\HH2 (\LH22-\HH4) \Big) 
\nonumber\\[.5ex]
&&{}
 - (\LH32-\HH5) \Big( \HH2  \LH32 +  \HH3\, \LH22 -\HH2\HH5 
\Big)
\nonumber\\[1.2ex]
&&{}
 +(\LH22-\HH4) (\LH22)^2  
\;,
\la{eq3introd}
\eea
as we shall show in the following.

Since the $\hcH_m$ themselves are combinations of the first order derivatives of the tau-function,
this equation is of the third order; it also has cubic non-linearity.
In the case $N=4$ the equation (\ref{eq3introd}) boils down to the standard Okamoto equation 
(\ref{Okameq}).

The paper is organized as follows. In Section 2 we derive the symmetrised form
 of the Schlesinger system.
In Section 3 we derive the generalized Okamoto equations. In Section 4 we show 
how the usual Okamoto equation is obtained from the generalized equation (\ref{eq3introd})
in the case $N=4$. In section 5 we discuss some open problems.

\section{Symmetrisation of Schlesinger system in terms of Virasoro generators}

\subsection{Variation of  a Riemann surface by vector fields
  on a contour}

The ``times'' $\lambda_j$ of the Schlesinger system can be viewed as
coordinates on the space of genus zero Riemann surfaces with $N$
punctures (we do not speak about the moduli space since we do not
identify two configurations of punctures related by a M\"obius
transformation).
The vectors $\p/\p\l_j$ span the tangent space to the space of
$N$-punctured spheres. However, there exist many different ways to
parametrize the tangent space to the space of Riemann surfaces of
given genus with a fixed number of punctures. 

One example is the variation of a Riemann surface by vector fields on
a chosen closed contour $l$ enclosing a disc $D$ (see
\cite{Konts}). To vary an unpunctured Riemann surface ${\cal L}$ of
genus $g$ by a vector field
${\bf v}$ on the contour  $l$ one cuts the disc $D$ and attaches it
back with infinitesimal shift of the boundary along the vector field
${\bf v}$. The complex structure of the Riemann surface also changes
infinitesimally (or remains the same). The infinite-dimensional space
$V$ of vector fields on $l$ can be represented as direct sum of three
subspaces,
$V_{\rm in}$ (which consists of vector fields which can be analytically
continued inside of the disc $D$), $V_{\rm out}$ (which consists of vector fields which can be analytically
continued outside of the disc $D$) and $V_0$ (which consists of vector
fields whose  analytical
continuation is impossible neither inside nor outside of the disc
$D$). The moduli of Riemann surface $\L$ are not changed by vector
fields from $V_{\rm in}$ and $V_{\rm out}$; the dimension of $V_0$ turns out to be equal to
$3g-3$ i.e.\ to dimension of the moduli space, and the vector fields
from $V_0$ do change the complex structure of the Riemann surface.

If the Riemann surface $\L$ has punctures (which are assumed to lie
outside of $D$), the vector fields from $V_{\rm out}$ should not only
admit analytical continuation to the exterior of $D$, but also vanish
at all the punctures.

To apply this general scheme in our present framework we choose the
contour $l$ to be a circle around $\lambda=\infty$ (such that all points
$\l_j$ lie ``outside'' of this circle). The standard basis in the space
of vector fields on $l$ is given by $v_m:= \lambda^{m+1} d/d\lambda$,
$m\in\Z$; the vectors $v_m$ satisfy standard commutation relations 
$[v_m, v_n]= (n-m)v_{m+n}$. How to relate variation of positions of
the punctures $\lambda_j$ to variation along vector fields $v_m$? Notice
that the fields $v_m$ with $m\leq -3$  can be holomorphically
continued in a neighbourhood of $\lambda=\infty$, and, therefore, do not
vary the singularities $\lambda_m$; they span the vector space
$V_{\rm in}$ in our case. The vector fields $v_m$ with $m=-2,
0, 1,\dots$ can not be analytically continued in the neighbourhood of
$\infty$. Being analytically continued in the neighbourhood of $0$,
these fields do not vanish at $\l_m$; however, there exists an
infinite-dimensional space (this is $V_{\rm out}$ in our case) of linear combinations of these $v_m$'s
which vanish at all points $\lambda_m$. The space $V_0$ can be chosen in
different ways. For example, $V_0$ can be chosen to be spanned by
all linear combinations of $v_{-1},\dots, v_{N-2}$. 

However, our goal will be to describe all Schlesinger equations
(independently of the number of singularities) within one
setting. Therefore, we shall vary $\lambda_j$ by all $v_m$ with
$m=-1,0,1,2,\dots$ in spite of the fact that for any given $N$ all these
vectors can be expressed as linear combinations of $N$ vectors $\p/\p\lambda_j$.

Namely, the action of the vector field $v_m$ on $\{\lambda_j\}$
  is given by the following linear combination of the tangent vectors $\p/\p\lambda_m$:
$$
L_m:=\sum_{j=1}^N \lambda_j^{m+1}\frac{\p}{\p\lambda_j}\;.
$$
 $L_m$ and $v_m$ coincide as tangent vector to the space of
$N$-punctured spheres. The vectors $L_m$ also satisfy commutation
relations of Virasoro algebra:
\be
[L_m,\;L_n]=(n-m) L_{m+n}\;.
\la{Vir}
\ee

\subsection{Schlesinger system in terms of Virasoro generators}

To symmetrise the Schlesinger equations we also  introduce the symmetric dependent variables:
\be
\cB_m := \sum_j \Gl_j^{m}\,A_j\equiv {\rm res}|_{\Gl=\infty}\{ \Gl^{m} A(\Gl)\}
\;.
\label{defB}
\ee
The new variable
\be
\cB_0 = \sum_j A_j 
\;,
\ee
plays a distinguished role: it vanishes on-shell (i.e.\ on solutions of the Schlesinger system);
however, off-shell it plays the role of a generator (with respect to the
Poisson bracket (\ref{PoissonA})) of constant gauge transformations
(i.e.\ constant simultaneous similarity transformations of all matrices $A_j$).

To describe the dynamics under the action of the differential operators $L_m$ we introduce the
symmetrised Hamiltonians $\cH_m$:
\be
\cH_m := -\frac{1}{2} {\rm res}|_{\Gl=\infty}{\rm tr}A^{2}(\Gl) \equiv \sum_j \Gl_j^{m+1}\,H_j
\;.
\la{Hamm}
\ee
These Hamiltonians can be expressed in terms of the variables  $\cB_k$ as follows:
\be
\cH_m = \widehat{\cH}_m 
+\ft14(m\!+\!1)\sum_j \Gl_j^m C_j
\;,
\la{HHmod}
\ee
where the modified Hamiltonians $\hcH$ are given by
\be
\hcH_m := -\ft14 \sum_{k=0}^m \tr\, \cB_k\,\cB_{m-k} 
\;,
\la{modham}
\ee
and $C_j:= \tr A_j^2 =\frac12 \alpha_j^2$.
In particular, 
$\hcH_{-1}=\hcH_{0}=\hcH_{1}=0$ (taking into account that
$\cB_{0}=0$), such that
the first three symmetrised Hamiltonians take the form
\be
\cH_{-1} = 0 \;,\qquad
\cH_{0}  = \ft14 \sum_j  C_j \;,\qquad
\cH_{1}  = \ft12 \sum_j \Gl_j C_j \;.
\ee

In terms of the Virasoro generators $L_m$ 
the equations (\ref{deftau}) for the Jimbo-Miwa $\tau$-function 
$\tau_{\rm JM}$ look as follows:
\be
L_m\,(\log \tau_{\rm JM})=\cH_m
\;.
\la{tauH}
\ee
It is convenient to introduce  also a modified $\tau$-function,
invariant under M\"obius transformations:
\begin{lemma}
The modified $\tau$-function $\tilde\tau$ defined by
\bea
\tilde\tau &\equiv& \tau_{\rm JM} \;\prod_{i<j} 
(\Gl_j - \Gl_i)^{-\ft1{2(N-2)} (C_i+C_j) + \ft2{(N-1)(N-2)} \cH_0}
\;,
\label{taumod}
\eea
is annihilated by the
first three Virasoro generators:
\be
L_{-1}\,\tilde\tau =L_{0}\,\tilde\tau =L_{1}\,\tilde\tau = 0 \;. 
\label{LLL101}
\ee
\end{lemma}
{\it Proof}: by straightforward computation.
\qed

In terms of the new variables~(\ref{defB}),
the Schlesinger system~(\ref{sch}) 
takes a very compact form:
\begin{theorem}
The differential operators $L_m$ act on the
symmetrised variables  $\cB_n$ as follows:
\be
L_m \cB_n =
\sum_{k=1}^{n-1} \left[\cB_k,\cB_{m+n-k} \right] + n \cB_{m+n} 
\;,
\la{Schmod}
\ee
for $m=-1,0,1,2,\dots$, $n=1,2,\dots$.
\end{theorem}

{\it Proof.} Using the Schlesinger equations (\ref{sch}), we have
$$L_m \cB_n \equiv
\sum_{i=1}^N\lambda_i^{m+1}\frac{\p}{\p\lambda_i}\left\{\sum_{j=1}^N
A_l\right\}
=
\sum_{i\neq j}\lambda_i^{m+1}
\frac{\lambda_i^n-\lambda_j^n}{\lambda_i-\lambda_j}[A_j,\,A_i]+
n\sum_{i=1}^N \lambda_i^{m+n} A_i
\;.
$$
Expanding
$$
\frac{\lambda_i^n-\lambda_j^n}{\lambda_i-\lambda_j}=\lambda_j^{n-1}+\lambda_j^{n-2}\lambda_i+\dots+
\lambda_j\lambda_i^{n-2}+\lambda_i^{n-1}\;,
$$
we further rewrite this expression for $L_m \cB_n$ as 
$$
[\cB_{n-1},\cB_{m+1}]+[\cB_{n-2},\cB_{m+2}]+\dots+[\cB_{1},\cB_{m+n+1}]+
n \cB_{m+n}\;,
$$
which coincides with the right hand side of (\ref{Schmod}).
 \qed
\hskip0.5cm

\begin{remark}
The system (\ref{Schmod}) can be equivalently rewritten as follows:
\be
L_m \cB_n =
\sum_{k=1}^{m} \left[\cB_k,\cB_{m+n-k} \right] + n \cB_{m+n} 
\;.
\la{Schmod1}
\ee
i.e.\ the right-hand side of (\ref{Schmod}) does not change if the
upper limit $n-1$ is substituted by $m$.
\end{remark}
The system of equations (\ref{Schmod}), or (\ref{Schmod1}) is the symmetric form of
the Schlesinger system.
Using (\ref{Schmod1}) we can express the commutators $[\cB_m,\cB_n]$ as follows:
\be
[\cB_m,\cB_n]= L_m\cB_n- L_{m-1}\cB_{n+1}+\cB_{m+n}
\;.
\ee
Acting on the modified hamiltonians $\hcH_n$ by the operators $L_m$, we get the following equation:
\be\la{LH}
L_m \hcH_n = -\ft12\,\sum_{k=1}^{n-1} k \,\tr \cB_{m+k}\cB_{n-k}\;.
\ee
In particular, we have
\be\la{LHcom}
L_m \hcH_n - L_n \hcH_m = (n-m)\,\hcH_{m+n}
\;.
\ee
The same equation holds for the Hamiltonians $\cH_n$ as a corollary of the
 integrability of equations (\ref{tauH}).

The Poisson bracket (\ref{PoissonA}) induces the following Poisson
 bracket between variables 
 $\cB_n$, $n=0,1,2,\dots$:
\be
\{\cB_n^a,\;\cB_m^b\}=f^{ab}{}_c\cB_{m+n}^c
\;.
\la{PoissonB}
\ee

Then equations (\ref{Schmod}) can then be written in the following
form:
\be
L_m \cB_n = \{\cH_m, \,\cB_n\} + n \cB_{m+n}\;.
\la{poissonform}
\ee
We note that formally the second term can be absorbed into the symplectic
action $\{\cH_m, \,\cB_n\}$ upon extending the affine Poisson structure (\ref{PoissonB})
by the standard central extension.


\section{Generalized Okamoto equations}


Here we shall use the symmetric form (\ref{Schmod}) of the Schlesinger
equations to derive an analog of Okamoto's equation (\ref{Okameq}) for
an arbitrary $2\times 2$ Schlesinger system.  In fact, one can write down a whole
family of scalar differential equations for the tau-function in terms
of the Virasoro generators $L_m$. In the next theorem we prove two
equations of this kind.

\begin{theorem}
The $\tau$-function $\tau_{\rm JM}$ (\ref{deftau})
 of an arbitrary $2\times 2$ Schlesinger system 
 satisfies the following two differential equations:
\begin{itemize}
\item
The third order equation  with cubic non-linearity:
\bea
\ft18\left(\LLH222 +2 \LH33 - 5 \LH42 -2 \HH6 \right)^2 
&=&
(\LH33-\LH42-\HH6) \, \Big( (\HH3)^2 +4\HH2 (\LH22-\HH4) \Big) 
\nonumber\\[.5ex]
&&{}
 - (\LH32-\HH5) \Big( \HH2  \LH32 +  \HH3\, \LH22 -\HH2\HH5 
\Big)
\nonumber\\[1.2ex]
&&{}
 +(\LH22-\HH4) (\LH22)^2  
\;.
\la{eq3}
\eea
\item
The fourth order  equation with quadratic non-linearity:
\bea
\LLLH2222 &=& 
8 \HH8 - 9 \LH44 + 10 \LH53 - 4 \LLH332 + 10 \LLH422  
\la{eq4}\\[1ex]
&&{}
-4 \Big(
\LH22 (2 \HH4 - 3 \LH22)  - \HH3 (\HH5 - 2 \LH32)  \Big)
 \nonumber\\[1ex]
 &&{}
-8 \HH2 \Big(2 \HH6 -3 \LH33+4 \LH42\Big) 
 \;,
 \nonumber
\eea
\end{itemize}
where according to (\ref{HHmod}), (\ref{tauH})
$$
\HH{m}= L_m\log \tau_{JM} -\f{m+1}{4}\sum_{j=1}^N \lambda_j^m C_j\;.
$$

\end{theorem}

{\it Proof.}
Inverting the system of equations~\Ref{LH}, we 
can express $\tr \,\cB_m\cB_n$ in terms of 
the Hamiltonians $\hcH_n$ as follows: 
\be
\tr\, \cB_m\cB_n = 4L_m\hcH_n 
-2\left( L_{m-1}\hcH_{n+1} + L_{m+1}\hcH_{n-1} \right)
\;.
\la{trBB}
\ee
>From the Schlesinger system (\ref{Schmod}), we furthermore get
\bea
L_k\,\tr(\cB_m\cB_n)&=&
\sum_{j=1}^k
\Big(
\tr (\cB_m[\cB_j,\cB_{n+k-j}]) +\tr (\cB_n[\cB_j,\cB_{m+k-j}])
\Big)
\nonumber\\ 
&&\quad {}+ n\,\tr\cB_m\cB_{k+n}+m\,\tr\cB_n\cB_{k+m}
\;.
\eea
Inverting this relation, we obtain for $k<m<n$
\bea
\tr \left( \cB_k [\cB_m,\cB_n] \right) &=&
\sum_{j=m}^{n-1} 
\Big(
L_k \,\tr \, \cB_j\cB_{m+n-j} 
- \ft12 L_{k-1} \, \tr \, \cB_{j+1} \cB_{m+n-j}
-\ft12 L_{k+1} \, \tr\, \cB_j \cB_{m+n-1-j}
\Big)
\nonumber\\[.5ex]
&&{}
+ \tr\, \cB_n\cB_{k+m} - \tr\, \cB_m\cB_{k+n}
\;.
\la{trBBB}
\eea
Combining this equation with (\ref{trBB}) we can thus express
also $\tr \left( \cB_k [\cB_m,\cB_n] \right)$ entirely in terms of
the action of the operators $L_m$ on 
the Hamiltonians $\hcH_{n}$, which can further be simplified upon using the
commutation relations (\ref{Vir}) and (\ref{LHcom}).
This leads to the closed expression
\bea
\tr \left( \cB_k [\cB_m,\cB_n] \right) &=&
2\,(L_{n-1}L_{m+1}-L_{n+1}L_{m-1})\,\hcH_k
\nonumber\\
&&{}
+2\,(L_{n+1}L_{m}-L_{n}L_{m+1})\,\hcH_{k-1}
\nonumber\\
&&{}
+2\,(L_{n}L_{m-1}-L_{n-1}L_{m})\,\hcH_{k+1}
\nonumber\\
&&{}
-4L_{m+n}\hcH_{k}+2L_{m+n+1}\hcH_{k-1}-2L_{m+n-1}\hcH_{k+1}
\nonumber\\
&&{}
-4L_{k+m}\hcH_{n}+2L_{k+m-1}\hcH_{n+1}+2L_{k+m+1}\hcH_{n-1}
\nonumber\\
&&{}
+4L_{k+n}\hcH_{m}-2L_{k+n-1}\hcH_{m+1}-2L_{k+n+1}\hcH_{m-1}
\;.
\label{trBBBfull}
\eea
In particular, for the lowest values of $k, m, n$ we obtain
\bea
\tr( \cB_1[\cB_2,\cB_3])
&=&
-2 L_2 L_2 \hcH_2
-4L_3\hcH_3 
+10 L_{4}\hcH_{2} 
+4\hcH_{6}
\;.
\label{trB123}
\eea
To derive from these relations the desired result, we make use of 
the following algebraic identity
\be\la{iden6}
{\rm tr}\left(M_1[M_2,M_3]\right)
{\rm tr}\left(M_4[M_5,M_6]\right)=-2
\Big({\rm tr}(M_1 M_4)\,{\rm tr}(M_2 M_5)\,{\rm tr}(M_3 M_6)+ ~\dots\Big)
\;,
\ee
valid for an arbitrary set of six matrices $M_j\in \mathfrak{sl}(2)$,
where the dots on the right-hand side 
denote complete antisymmetrisation of the expression 
with respect to the indices $1, 2, 3$.
In terms of the structure constants of $\mathfrak{sl}(2)$,
this identity reads
\bea
f_{abc} f^{def} &=& 
6 \,\delta_{[a}^{[d}\,\delta_{\vphantom{[}b}^{\vphantom{[}e}\, \delta_{c]}^{f]}
\;,
\label{fdelta}
\eea
where adjoint indices $a, b, \dots$
are raised and lowered with the Cartan-Killing form.
Setting in (\ref{iden6}) 
$M_1=M_4=\cB_1$, $M_2=M_5=\cB_2$  and $M_3=M_6=\cB_3$, and using (\ref{trBB}), (\ref{trB123}), we arrive (after some calculation) at (\ref{eq3}).
\medskip

Equation~(\ref{eq4}) descends from another
algebraic identity  
\bea\la{iden4}
{\rm tr}\left([M_1,M_2][M_3,M_4]\right)&=& 
-2\,\Big({\rm tr}(M_1 M_3)\,{\rm tr}(M_2 M_4)-
{\rm tr}(M_2 M_3)\,{\rm tr}(M_1 M_4)\Big)
\;,
\eea
valid for any four $\mathfrak{sl}(2)$-valued matrices $M_j$.
In terms of the structure constants of $\mathfrak{sl}(2)$,
this identity reads
\bea
f_{abf} f^{cdf} &=& 
2 \,\delta_{[a}^{[c}\, \delta_{b]}^{d]}
\;,
\label{fdelta2}
\eea
and is obtained by contraction from (\ref{fdelta}).
We consider the action of $L_2$ on (\ref{trB123}) which yields
\bea
3\tr( \cB_1[\cB_2,\cB_5])-
2\tr( \cB_1[\cB_3,\cB_4])&=&
\tr[\cB_1,\cB_3][\cB_1,\cB_3] 
-\tr[\cB_1,\cB_2][\cB_1,\cB_4] 
-\tr[\cB_1,\cB_2][\cB_2,\cB_3]
\nonumber\\[.5ex]
&&{} 
-2 L_2L_2 L_2 \hcH_2
-4L_2L_3\hcH_3 
+10 L_2L_{4}\hcH_{2} 
+4L_2\hcH_{6}
\;.
\nonumber
\eea
The l.h.s.\ of this equation can be reduced by (\ref{trBBB})
while the first terms on the r.h.s.\ are reduced by means 
of the algebraic relations (\ref{iden4}) together with (\ref{trBB}).
As a result we obtain equation (\ref{eq4}).
\qed


\mathon
\section{Four simple poles: reproducing the Okamoto equation}
\mathoff


As remarked above, the explicit form of the differential equations 
(\ref{eq3}), (\ref{eq4}) for the $\tau$-function $\tau_{\rm JM}$ is obtained
upon expressing the modified Hamiltonians~$\HH{m}$ in terms of $\tau_{\rm JM}$
by virtue of (\ref{HHmod}), (\ref{tauH}).
As an illustration, we will work out these equations for the 
Schlesinger system with four singularities and show that they reproduce precisely Okamoto's equation~(\ref{Okameq}). For $N=4$, the modified $\tau$-function
$\tilde\tau$ from (\ref{taumod}) depends only on the cross-ratio
\bea
t&=&
\frac{(\lambda_1-\lambda_3)(\lambda_2-\lambda_4)}{(\lambda_1-\lambda_4)(\lambda_2-\lambda_3)}
\;.
\eea
We furthermore define the auxiliary function
\bea
S(t)&=&2t(1+t)\f{d}{dt}\log\tilde\tau(t)
\;.
\eea
Then equation (\ref{eq3}) in terms of $S$ 
after lengthy but straightforward calculation gives rise to
the second order differential equation
\bea
3\left(t(1-t) \,S''\right)^2  &=&
3( C_2 - C_3)( C_1 - C_4)\,(S-tS')
-3(C_1 - C_3)( C_2 - C_4)\,S'
\nonumber\\[.5ex]
&&{}
+12\,(S-tS') S'{}^2
+12 (S-tS')^2 S'
\nonumber\\
&&{}
+2\s_1[C]\,\Big(
(S-tS')^2
+ (S-tS')S'
+ S'{}^2
\Big)
\nonumber\\
&&{}
-\ft1{18}\s_1[C]^3+\ft12 \s_1[C] \, \s_2[C]  -3\s_3[C]
\;.
\la{eqS}
\eea
where $\s_i[C]$ are the elementary symmetric polynomials of the $C_i$'s
$$
\sigma_1[C]=C_1+ C_2+C_3+C_4\;,\qquad
\sigma_2[C]= \sum_{j<k} C_j C_k\;,\qquad 
\s_3[C]= \sum_{j<k<l} C_j C_k C_l\;.
$$
Finally, 
it is straightforward to verify that
with $h(t)\equiv S(t) - \f{1}{12}(1-2t)\,\s_1[C]$, 
equation (\ref{eqS}) is equivalent to Okamoto's equation (\ref{Okameq}).
\bigskip

In turn, equation (\ref{eq4}) leads to the following quadratic,
third order differential equation in the function $S$:
\begin{eqnarray}
6t( 1 - t) \left( ( 1 - 2t) \,S'' +
     t( 1 - t) \,S^{(3)} \right)   
     &=&   3\,\Big( \left( C_1 - C_3 \right) \left( C_2 - C_4 \right)  -
      \left( C_2 - C_3 \right) \left( C_1 - C_4 \right) t \Big)  
      \nonumber\\
&&-
   12\,{{S}^2} +2( 1 - 2t ) \,S\,
    \left( \s_1[C] - 12\,S' \right)  \nonumber\\[.5ex]
&&+
   4 \s_1[C] \,
   ( 1 - t + {t^2} ) \,S' +
   36t( 1 - t ) \,{{S'}^2}\;.
   \nonumber\\
\la{thirdorder}
\end{eqnarray}
Indeed, this equation can
also be obtained by straightforward differentiation of (\ref{eqS}) with respect to $t$.
In terms of the function $h$ equation (\ref{thirdorder}) takes the following form
\bea
(1-t)t\Big(6 h'{}^2-(1-2t) h''-(1-t)t \,h^{(3)}\Big)
&=&
4 h^2 +8 (1-2t)\, h h'
-2(b_1^2+b_2^2+b_3^2+b_4^2)\,h'
\nonumber\\[.5ex]
&&{}
-\prod_{i<j} b_i^2 b_j^2 +2(1-2t)\, b_1b_2b_3b_4
\;,
\la{thirdorder1}
\eea
equivalently obtained by derivative of Okamoto's equation~(\ref{Okameq}).


\mathon
\section{Discussion and Outlook}
\mathoff


We have shown in this paper that the symmetric form
(\ref{Schmod}), (\ref{Schmod1}) of the Schlesinger system
gives rise to a straightforward algorithm that allows to
translate the algebraic $\mathfrak{sl}(2)$ identities 
(\ref{fdelta}), (\ref{fdelta2}) into differential equations
for the $\tau$-function of the Schlesinger system.
In the simplest case of four singularities, the
resulting equations reproduce the known
Okamoto equation~(\ref{Okameq}).
In the case of more singularities, the same equations (\ref{eq3}), (\ref{eq4})
give rise to a number of non-trivial differential equations
to be satisfied by the $\tau$-function.

Apart from this direct extension of Okamoto's equation,
the link between the algebraic structure of $\mathfrak{sl}(2)$
and the Schlesinger system's $\tau$-function gives rise to 
further generalizations.
Note, that in the proof of Theorem~2,
with equation~(\ref{trBBBfull}) we have already given
the analogue of (\ref{trB123}) to arbitrary values of $k, m, n$.
Combining this equation with the identity (\ref{iden6}) thus gives
rise to an entire hierarchy of third order equations that generalize (\ref{eq3}).
Likewise, the construction leading to the fourth order equation (\ref{eq4})
can be generalized straightforwardly upon applying (\ref{iden4})
to other Virasoro descendants of the cubic equation.

As an illustration, we give the first three equations of the hierarchy 
generalizing (\ref{eq4}):
\ba
\LLLH3222 
&=&
6 \HH9 - 6 \LH54 + 10 \LH63 - 5 \LH72 - \LLH333+ 
    \LLH432 + 6 \LLH522
\non[1ex]
&&
{}+ 
8 \HH2(2\LH43\!-\!3\LH52\!-\!\HH7) + 4\HH3(\LH33\!-\!\LH42)
- 8 \HH4\LH32 + 12 \LH22\LH32 
\;,
\nonumber\\[3ex]
\LLLH4222 
&=&
2 \HH{10}+ 3 \LH55- 8 \LH64 + 9 \LH73 - 9 \LH82 
- \LLH433
\non[1ex]
&&
{} + 4 \LLH442
- 3 \LLH532 + 6 \LLH622
{}+8\HH2( \LH44\!+\!3 \LH53\!-\!3 \LH62 ) 
\non[1ex]
&&
-4 \HH3 (\LH43\!-\!3 \LH52)
+8 \HH4 \LH42 - 12 \LH22 \LH42 
\;,
\nonumber\\[3ex]
\LLLH3322
&=&
8 \HH{10} - 6 \LH55 +4 \LH64 + 6 \LH73 - 6 \LH82 
{}-
 \LLH433 - 2 \LLH442 + 6 \LLH532
\non[1ex]
&&
{}
 + 4 \LLH622 
 - 4(\HH5 - 2 \LH32) \LH32 +4 \LH22 \LH33 -8 \HH4 \LH42 
\non[1ex]
&&
+4 \HH3 (\LH43\!-\!3\LH52) 
-8 \HH2 \left(2 \HH8\!-\!2\LH44\!+\!\LH53\!+\!2\LH62\right)
\;.
\ea

Obviously, these equations are not all independent, but
related by the action of the lowest Virasoro generators
$L_{\pm1}$, using that
\bea
L_1 (\LLLH2222 ) &=& 
4\, (\LLLH3222 ) + \dots 
\;,
\eea
etc., since $\HH1=0$.
The number and structure of the independent equations in this hierarchy
is thus organized by the structure of representations of the Virasoro algebra.
For the case of $N=4$ singularities, the explicit form of all the
equations of the hierarchy reduces to equivalent forms of 
(\ref{Okameq}) and (\ref{thirdorder1}).
With growing number of simple poles, the number of independent 
differential equations induced by the hierarchy increases.

Therefore, we arrive to a natural question: which set of derived
equations for the tau-function is equivalent to the original
Schlesinger system? We stress that all differential equations for the
tau-function are PDE with respect to the variables
$\l_1,\dots,\l_N$. However, if one gets a sufficiently high number of
independent equations, one can actually come to a set of ODE's for the
tau-function. This situation resembles the situation with the original
form of the Schlesinger system (\ref{sch}): if one ignores the second
set of equations in (\ref{sch}), one gets a system of PDE's for the
residues $A_j$; only upon adding the equations for $\p A_j/\p\l_j$ one gets
a system of ODE's with respect to each $\l_j$ (the flows with respect
to different $\l_j$   commute).

Let us finally note that the construction we have presented in
order to derive the differential equations (\ref{eq3}), (\ref{eq4})
suggests a number of interesting further generalizations that deserve
further study.

\begin{itemize}

\item
At the origin of our derivation have figured the
algebraic $\mathfrak{sl}(2)$ identities (\ref{fdelta}), (\ref{fdelta2})
that we have translated into differential equations.
Similar identities exist also for higher rank groups
(e.g.\ $M>2$, or the Schlesinger system for
orthogonal, symplectic, and exceptional groups)
where the number of independent tensors may be larger. 
It would be highly interesting to understand if equations analogous 
to (\ref{eq3}), (\ref{eq4}) can be derived from such higher rank algebraic identities.
As those identities will be built from a larger number of 
invariant tensors (structure constants, etc.),
the corresponding differential 
equations would be of higher order in derivatives.

\item
Is it possible to combine our present construction applicable to
Schlesinger systems with simple poles only with construction of
\cite{Harnad} which requires the presence of higher order poles? 
What would be the full set of equations for the tau-function with
respect to the full set of deformation parameters in presence of
higher order poles?

\item
The Schlesinger system (\ref{sch}) has also been constructed for
various higher genus Riemann surfaces~\cite{KS,Kawai,LO,Taka}.
It would be interesting to first of all find the proper
generalization of the symmetric form (\ref{Schmod}), (\ref{Schmod1}) 
of the Schlesinger system to higher genus surfaces
which in turn should allow to derive by an analogous construction
the non-trivial differential equations satisfied by the
associated $\tau$-function. We conjecture that in some sense the form
(\ref{Schmod}) should be universal: it should remain the same,
although the definition of the Virasoro generators $L_m$ and the
variables $\cB_m$ may change.

\item
As we have mentioned above, 
the extra term $n \cB_{m+n}$ in the Hamiltonian dynamics of the 
symmetrised Schlesinger system (\ref{poissonform})
can be absorbed into the symplectic action upon replacing the
standard affine Lie-Poisson bracket (\ref{PoissonB}) by its centrally extended version. 
However, this central extension is not seen in any
of the finite-$N$ Schlesinger systems. This seems to suggest that the system (\ref{Schmod})
should be considered not just as a symmetric form of the usual Schlesinger system with finite number of poles, but as a 
``universal'' Schlesinger system which involves an infinite set of independent variables
$\cB_n$.  Presumably, this full system involves the generators $L_n$ and coefficients $\cB_n$ not only for
positive, but also for negative $n$.

In this setting, the centrally extended version of the bracket (\ref{PoissonB}) should appear naturally.
The most interesting problem would be to find the geometric origin of such a generalized system; 
a possible candidate could be the isomonodromic deformations on higher genus curves.

\end{itemize}

\bigskip
\bigskip

\noindent
{\bf Acknowledgements:}
The work of H.S.\ is supported in part by the Agence Nationale de la Recherche (ANR).
The work of D.K. was supported by NSERC, NATEQ and Concordia
University Research Chair grant.


\begingroup\raggedright\endgroup

\end{document}